\documentclass[11pt]{article}
\usepackage{vietnam,epsfig}

\def\Journal#1#2#3#4{{#1} {\bf #2}, #3 (#4)}

\def\AA{\em A\&A}
\def\APJ{\em ApJ}
\def\APJS{\em ApJS}
\def\SSR{\em Space Sc. Rev.}

\def\be{\begin{equation}}
\def\ee{\end{equation}}
\def\bea{\begin{eqnarray}}
\def\eea{\end{eqnarray}}

\begin{document}
\vspace*{4cm}
\title{PHYSICAL IMPLICATIONS OF INTEGRAL/SPI GAMMA-RAY LINE MEASUREMENTS OF
THE 2003 OCTOBER 28 SOLAR FLARE}

\author{V. Tatischeff$^1$, J. Kiener$^1$ and M. Gros$^2$}

\address{$^1$CSNSM, IN2P3-CNRS and Universit\'e Paris-Sud, 91405 Orsay Campus,
France\\
$^2$DSM/DAPNIA/SAp, CEA Saclay, 91191 Gif-sur-Yvette, France}

\maketitle\abstracts{
The very powerful X-class solar flare of 2003 October 28 was detected with the
INTEGRAL spectrometer as an intense $\gamma$-ray flash of about 15 minutes.
Despite the non-standard incidence of the solar $\gamma$-rays, time-resolved 
spectra including several nuclear $\gamma$-ray lines were obtained. Such a
measurement with a high-energy-resolution instrument can provide valuable 
information of the isotopic abundances of the ambient solar material, as well 
as the composition, directionality and energy spectra of the accelerated 
nuclei. First results on the measured $\gamma$-ray line ratios and time
history of the neutron-capture line are presented.}

\section{Introduction}
Although not devoted to solar physics, the INTEGRAL observatory can
provide valuable measurements of the high-energy emission produced in the
most powerful solar flares. In particular, the spectrometer SPI~\cite{ve} can
allow fine spectroscopic analyses of the nuclear $\gamma$-ray line emission. 
These lines are produced by the interactions of flare-accelerated particles 
with the ambient solar material and their intensities can be used to 
determine isotopic abundances in the solar atmosphere~\cite{ra}. In addition, 
measurements of the shape and redshift of some deexcitation lines can provide 
complementary information on the directionality and spectra of the 
accelerated ions~\cite{sm}. 

Very high solar activity was observed in late October and early November 
2003, with several flares of class X originating from the same active region 
of the Sun. In this paper, we briefly review the data obtained with SPI for 
the X17.2 flare of October 28~\cite{gr}. We then present a preliminary 
analysis of the measured $\gamma$-ray line fluence ratios and of the time 
history of the 2.22~MeV neutron-capture line. 

\section{Observations}

The $\gamma$-ray emission from the 2003 Oct. 28 flare was detected with SPI
as an intense flash of $\sim$15 minutes starting at 11:02 UT. The satellite
was then observing the supernova remnant IC443, with the Sun at 122$^\circ$
from the instrument line-of-sight. Background-subtracted spectra were
obtained by using both single and multiple events from the 19 Ge
detectors. Four deexcitation $\gamma$-ray lines produced by the interactions 
of flare-accelerated protons and $\alpha$-particles with the solar
atmosphere were detected: at 4.44 MeV from ambient $^{12}$C*, and 
6.13, 6.92 and 7.12 MeV from ambient $^{16}$O*. In addition, we measured a
strong line emission at 2.22 MeV from radiative capture of secondary
neutrons by photospheric H. The relative fluences of these lines are given
in Table~1. They were obtained from preliminary simulations of $\gamma$-ray 
transmission through the satellite material, for the configuration of 
INTEGRAL during the flare. 

\begin{table}[t]
\caption{Relative line fluences for the 2003 October 28 flare}
\vspace{0.4cm}
\begin{center}
\begin{tabular}{|c|c|}
\hline
Line Energy (MeV) & Relative Fluence \\ 
\hline
2.22 & 10.7$\pm$1.7 \\
4.44 & 0.92$\pm$0.14 \\
6.13 & $\equiv$1.00$\pm$0.17 \\
6.92 & 0.33$\pm$0.13 \\
7.12 & 0.20$\pm$0.12 \\
\hline
\end{tabular}
\end{center}
\end{table}

\section{Line fluence ratios}

\begin{figure}[b!]
\centering
\psfig{figure=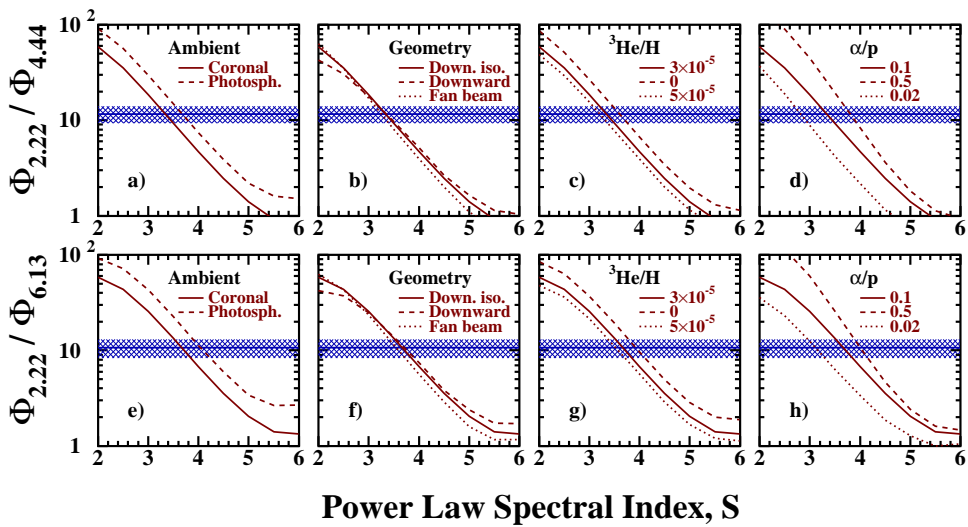,width=0.81\linewidth}
\caption{Neutron capture line to $^{12}$C and $^{16}$O line fluence ratios.
Horizontal hatched areas: SPI data. Solid curves show calculated ratios as a
function of power-law spectral index, for the following case: accelerated 
ions with $\alpha$/$p$=0.1 and downward-isotropic distribution, 
interacting with an ambient medium of coronal composition and
$^3$He/H=3$\times$10$^{-5}$. Dashed and dotted curves show the effects of 
changing the ambient medium composition (panels {\it a} and {\it e}) the 
flare geometry ({\it b} and {\it f}) the ambient $^3$He abundance ({\it c} 
and {\it g}) or the $\alpha$/$p$ ratio ({\it d} and {\it h}).}
\end{figure}

Calculations of $\gamma$-ray line emission were performed assuming a 
thick target interaction model and a power-law source spectrum for the
accelerated particles~\cite{ra}. We considered both photospheric~\cite{lo} and
coronal~\cite{re} compositions for the ambient medium. For the production of
the neutron-capture 2.22 MeV line, we used the code developed by Hua et 
al~\cite{hu}. Whereas the emission of the $^{12}$C and $^{16}$O narrow 
lines is only due to accelerated proton and $\alpha$-particle interactions, 
neutron production also depends on the fast heavier nuclei. We 
assumed for these particles an impulsive-flare average composition obtained 
from measurements of solar energetic particle events~\cite{re}, but allowed 
the $\alpha$/$p$ abundance ratio to vary between 0.02 and 0.5. The cross 
sections for the production of the $^{12}$C and $^{16}$O lines are based on 
Kozlovsky et al~\cite{ko}. 

The neutron capture-to-deexcitation line fluence ratios strongly depend on
the accelerated particle spectral index $S$, because, on average, the neutrons
are produced at higher energies than the $^{12}$C and $^{16}$O lines (see 
Fig.~1). But these ratios also depend on the composition of the 
accelerated particles and the ambient material, as well as on the angular 
distribution of the interacting ions. We used the calculations shown in 
Fig.~1 to take into account these uncertainties and obtained from the 
comparaison with SPI data: $S$=3.15$\pm$0.55, 3.65$\pm$0.55 and 
4.075$\pm$0.475, for $\alpha$/$p$=0.02, 0.1 and 0.5, respectively. 

\begin{figure}[t!]
\centering
\psfig{figure=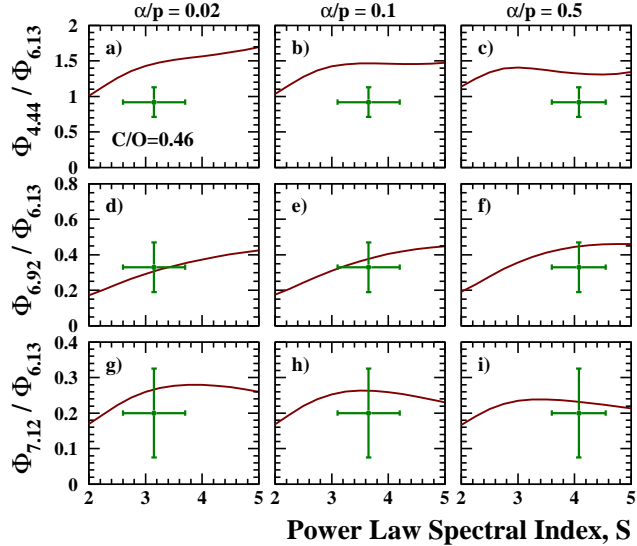,width=0.53\linewidth}
\caption{Observed (crosses) and calculated (solid lines) $^{12}$C (4.44 MeV) 
and $^{16}$O (6.13, 6.92 and 7.12 MeV) line flux ratios as a function of 
{\it S}, for 3 values of $\alpha$/$p$: 0.02 (panels {\it a}, 
{\it d} and {\it g}), 0.1 ({\it b}, {\it e} and {\it h}) and 0.5 
({\it c}, {\it f} and {\it i}).}
\end{figure}

In Fig.~2, we compare calculated $^{12}$C and $^{16}$O line 
ratios for these 3 values of $\alpha$/$p$ with the SPI data for the 
corresponding values of $S$. We assumed that the $\gamma$-ray production 
region is of coronal composition, with C/O=0.46~\cite{re}. We see that the 
calculated $^{16}$O line ratios are in good agreement with the SPI data, 
given the large uncertainties due to the relatively low statistics obtained 
for the $\sim$7~MeV lines. But we also see that the calculated 4.44-to-6.13 
MeV line ratios overestimate the observed value, for all the 
$\alpha$/$p$ ratios. We have further compared our calculations with data 
obtained for other solar flares with {\it SMM}, CGRO/{\it OSSE} and 
{\it RHESSI} and found the theoretical $^{12}$C-to-$^{16}$O line ratios to 
overestimate the average of the measured flux ratios by a factor of 
$\sim$1.5~\cite{ta}. This discrepancy suggests that the $^{12}$C and/or 
$^{16}$O abundances at the $\gamma$-ray production sites could be
different from their generally-assumed coronal values. 

\section{Time dependence of the 2.22 MeV line emission}

Time history measurements of the 2.22 MeV line can provide a unique
determination of the photospheric $^3$He/H ratio, because neutron capture on
$^3$He via the $^3$He($n$,$p$)$^3$H reaction can significantly shorten the
delay of the radiative-capture line emission. But as recently emphasized 
by Murphy et al.~\cite{mu}, the time history of this line is also strongly
dependent on the angular distribution of the interacting flare-accelerated 
particles. Following these authors, we calculated the 2.22 MeV line
production by using a detailed magnetic loop model~\cite{hu}, for which this
angular distribution is parameterized by the level of MHD pitch-angle
scattering (PAS) occuring in the coronal portion of the loop. We assumed the
neutron-production time history to be identical to the one of the prompt
deexcitation $\gamma$-ray line emission, for which good quality data were
obtained with SPI (Fig.~3a). 

Fig.~3b shows the best fit to the measured count rate of the 2.22 MeV line 
and Fig.~4 the regions of 1$\sigma$ and 90\% confidence levels for the two
free parameters. These results are in good agreement with those previously 
obtained with {\it RHESSI}~\cite{mu}. However, the derived $^3$He/H ratio is 
still not well constrained. But the 4.44 and 6.13~MeV line shapes measured
with SPI will allow an independent determination of $\lambda$, which should
significantly reduce the uncertainty on the photospheric $^3$He abundance
(Fig.~4). 

\begin{figure}[t!]
\centering
\psfig{figure=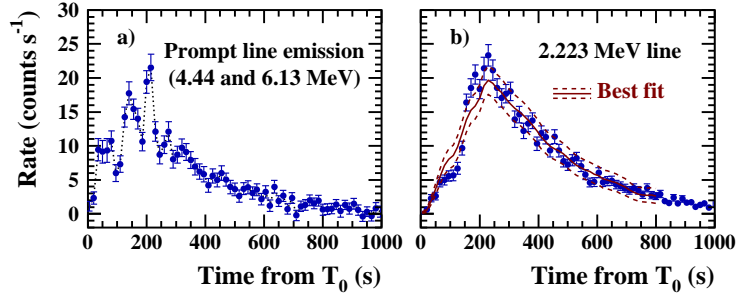,width=0.62\linewidth}
\caption{Measured time dependences of ({\it a}) the sum of the 4.44 and 6.13 
MeV line count rates and ({\it b}) the 2.22 MeV neutron-capture line 
emission. Also shown in panel ({\it b}) is the best fit with $\pm$1$\sigma$ 
uncertainties, obtained for the normalized PAS mean free path 
$\lambda$=2000 and for $^3$He/H=6$\times$10$^{-5}$ (see text).}
\end{figure}

\begin{figure}[t!]
\centering
\psfig{figure=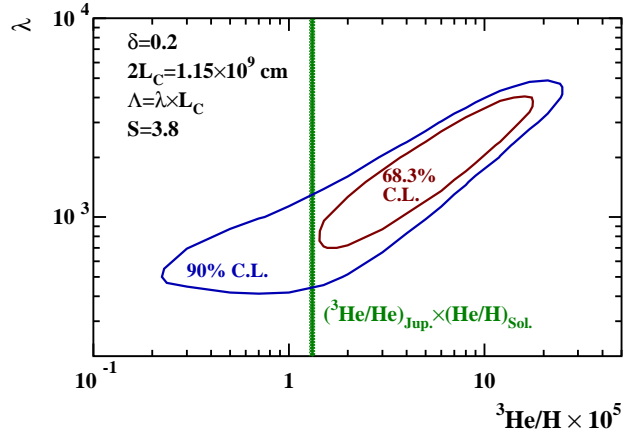,width=0.52\linewidth}
\caption{Regions of 68.3\% (1$\sigma$) and 90\% confidence levels for the 
two free parameters $\lambda$ and $^3$He/H. Also shown is the estimated solar 
$^3$He abundance obtained by multiplying the He/H ratio measured from 
heliosismology with the $^3$He/$^4$He ratio as determined for Jupiter's 
atmosphere.}
\end{figure}

\section*{Acknowledgments}
This work is based on observations with INTEGRAL, an ESA project with 
instruments and science data centre funded by ESA member states (especially 
the PI countries: Denmark, France, Germany, Italy, Switzerland, Spain), Czech 
Republic, Poland, Russia and the USA. We acknowledge A. Bykov for permission 
to use the data prior to their public release.

\end{document}